\documentclass[aip,jap,amsmath,amssymb,reprint]{revtex4-1}

\usepackage{graphicx}
\usepackage{dcolumn}
\usepackage{units}
\usepackage{hyperref}

\newcommand{\tsub}[1]{\textnormal{\tiny{#1}}}

\begin{document}

\title[]{Precise determination of micromotion for trapped-ion optical clocks}

\author{J. Keller}
\affiliation{Physikalisch-Technische Bundesanstalt, Bundesallee 100, 38116 Braunschweig, Germany}
\author{H.~L. Partner}
\affiliation{Physikalisch-Technische Bundesanstalt, Bundesallee 100, 38116 Braunschweig, Germany}
\affiliation{Present Address: Institut f\"ur Physik, Humboldt-Universit\"at zu Berlin, Newtonstra{\ss}e 15, 12489 Berlin, Germany}
\author{T. Burgermeister}
\affiliation{Physikalisch-Technische Bundesanstalt, Bundesallee 100, 38116 Braunschweig, Germany}
\author{T.~E. Mehlst\"aubler}
\affiliation{Physikalisch-Technische Bundesanstalt, Bundesallee 100, 38116 Braunschweig, Germany}
\email{tanja.mehlstaeubler@ptb.de}

\date{\today}

\begin{abstract}
  As relative systematic frequency uncertainties in trapped-ion spectroscopy are approaching the low $10^{-18}$ range, motional frequency shifts account for a considerable fraction of the uncertainty budget. Micromotion, a driven motion fundamentally connected to the principle of the Paul trap, is a particular concern in these systems. In this article, we experimentally investigate at this level three common methods for minimizing and determining the micromotion amplitude. We develop a generalized model for a quantitative application of the photon-correlation technique, which is applicable in the commonly encountered regime where the transition linewidth is comparable to the rf drive frequency. We show that a fractional frequency uncertainty due to the 2nd-order Doppler shift below \mbox{$\vert\Delta\nu/\nu\vert=1\times10^{-20}$} can be achieved. The quantitative evaluation is verified in an interleaved measurement with the conceptually simpler resolved sideband method. If not performed deep within the Lamb-Dicke regime, a temperature-dependent offset at the level of $10^{-19}$ is observed in resolved sideband measurements due to sampling of intrinsic micromotion. By direct comparison with photon-correlation measurements, we show that the simple to implement parametric heating method is sensitive to micromotion at the level of \mbox{$\vert\Delta\nu/\nu\vert=1\times10^{-20}$} as well.
\end{abstract}

\maketitle
\section{Introduction}
Laser cooled ions in Paul traps \cite{Paul1990} are excellent candidates for precision spectroscopy, \cite{Chou2011,Huntemann2012,Barwood2014} quantum information processing \cite{Cirac1995,Blatt2008} and quantum simulation. \cite{Porras2004,Friedenauer2008,Islam2013} High trap depths allow trapping times ranging from hours to months and the strong confinement permits localization to within a few nanometers, allowing for precise control of energy shifts due to external fields. With trapped ion frequency standards approaching fractional frequency uncertainties below $10^{-17}$, motional shifts are currently among the most significant contributions. \cite{Chou2010,Huntemann2012} Of particular concern is the 2nd-order Doppler shift arising from micromotion, a periodic motion driven by the confining radio-frequency (rf) electric field. Methods for the precise determination of the micromotion amplitude are therefore required in order to minimize its origins and to characterize the residual uncertainty. A similar level of control over micromotion is required in experiments studying the interaction between cold trapped ions and ensembles of neutral atoms, \cite{Grier2009,Schmid2010,Zipkes2010,Hall2011,Rellergert2011} as its kinetic energy might otherwise dominate the dynamics.\\

The origins and consequences of micromotion as well as methods for its determination have been discussed extensively by Berkeland et al. \cite{Berkeland1998} Various other techniques, suitable under different experimental conditions, have been investigated since then. \cite{Lisowski2005,Brown2007,Allcock2010,Ibaraki2011,Chuah2013,Haerter2013,Gloger2015} The decreasing overall frequency uncertainty in trapped-ion spectroscopy prompts for an increasingly detailed understanding of the uncertainty in micromotion detection. In this article, we experimentally investigate the reliability of three of the most commonly used methods, namely sideband spectroscopy, the photon-correlation method, and parametric excitation. We show that each is capable of achieving a fractional frequency uncertainty due to the 2nd-order Doppler shift on the order of $10^{-20}$, independent of the mass of the ion. A new model is developed for photon-correlation signals, which is necessary for a quantitative evaluation under commonly encountered experimental conditions. It is verified by comparison with resolved sideband measurements. We derive an expression for a temperature-dependent offset seen in resolved sideband measurements due to intrinsic micromotion if the ion is not deep within the Lamb-Dicke regime. This contribution becomes relevant at the level of $10^{-19}$ for our parameters. We also discuss the applicability of each method during the operation of an optical clock, where the disruption of the interrogation needs to be minimized.\\

This paper is organized as follows: Section \ref{ionmotion} reviews ion motion in a linear Paul trap, the origins of micromotion and the resulting frequency shifts. Section \ref{emm_section} introduces the three evaluated techniques for minimizing excess micromotion and experimentally compares their resolution limits and suitability for clock operation. The results are summarized in section \ref{summary}.

\section{\label{ionmotion}Motion of trapped ions}
In the following, we review the motion of a single ion in a Paul trap. For brevity, we restrict the discussion to linear traps, as this geometry is used in the experiments presented here. In a linear trap, rf confinement is only used for the two radial dimensions, while the axial trapping potential is provided by a static electric field. Our results can be directly extended to Paul traps with three-dimensional rf confinement.\\

\begin{figure}
  \centerline{\includegraphics[width=.45\textwidth]{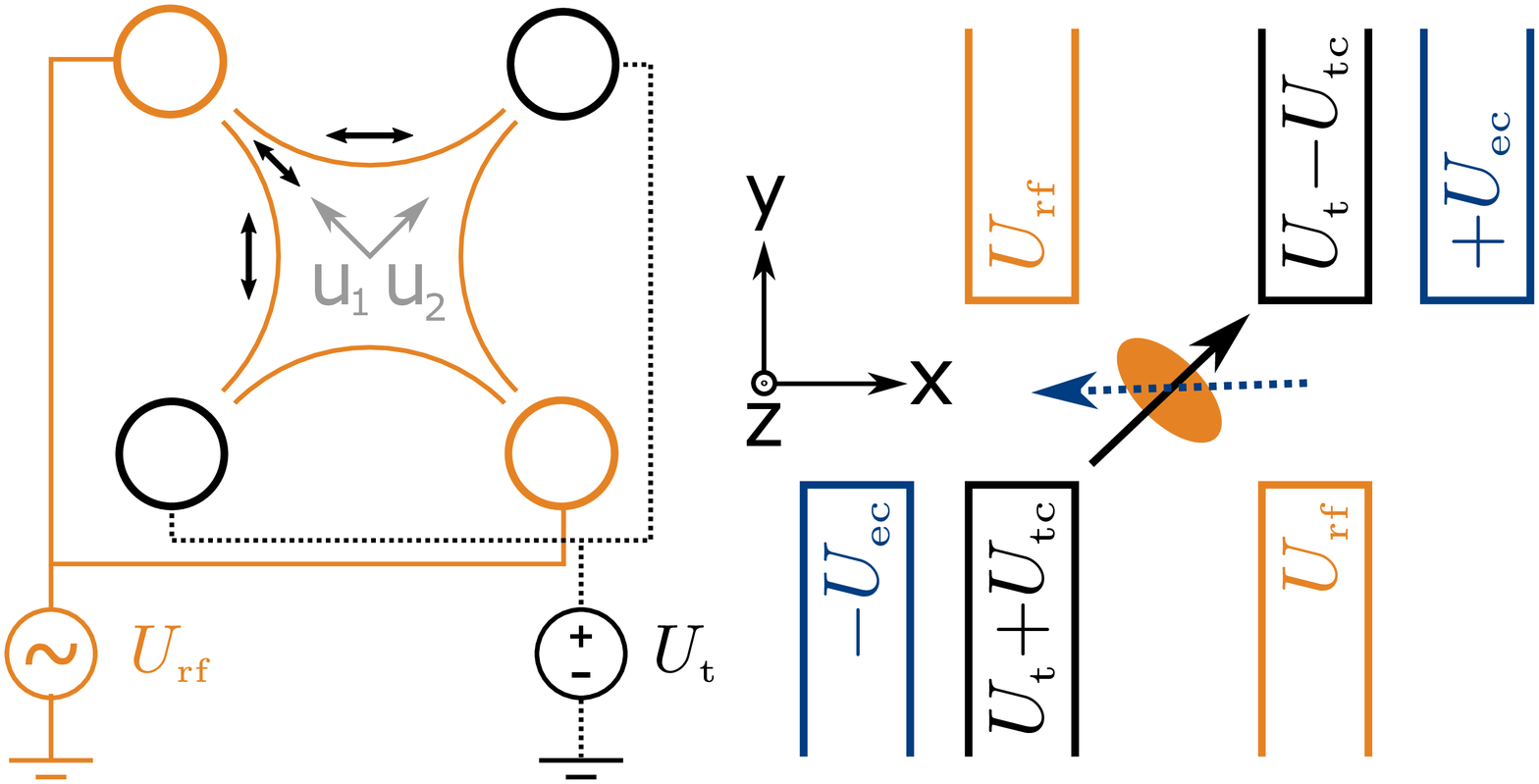}}
  \caption{\label{trap_schematic} Left: radial plane of a linear Paul trap with asymmetric rf drive and applied dc voltage $U_\tsub{t}$ to remove the degeneracy of radial secular frequencies. Black arrows indicate the direction of the micromotion, which depends on the rf electric field lines at the ion position. Right: the electrode geometry of the trap used in the experiments presented here. The compensation voltages $U_\tsub{tc}$ and $U_\tsub{ec}$ control the ion position in the radial plane, as indicated by the arrows.}
\end{figure}

Without loss of generality, we assume that the principal axes of the radial trapping potential are aligned along the directions $u_{1}$ and $u_{2}$, as shown in Fig.~\ref{trap_schematic}, to obtain decoupled equations of motion. They are inhomogeneous forms of the Mathieu equation \cite{Paul1990}
\begin{equation}
  \ddot{u}_i+\left(a_i+2q_i\cos(\Omega_\tsub{rf}t)\right)\frac{\Omega_\tsub{rf}^2}{4}u_i=\frac{QE_{\tsub{dc},i}}{m}\;\textnormal{,}
  \label{Mathieu}
\end{equation}
where $Q$ and $m$ are the charge and mass of the ion, $\Omega_\tsub{rf}$ is the trap drive frequency, $E_{\tsub{dc},i}$ is the component of static electric field along $u_i$, and the $q_i$ and $a_i$ parameters are related to the applied voltages. They are defined as
\begin{equation}
  q_1=-q_2=\frac{2Q\kappa_\tsub{rf}U_\tsub{rf}}{md^2\Omega_\tsub{rf}^2}\quad\textnormal{and}
\end{equation}
\begin{equation}
  a_{1,2}=-\frac{2\omega_\tsub{z}^2}{\Omega_\tsub{rf}^2}\pm\frac{4Q\kappa_\tsub{t}U_\tsub{t}}{md^2\Omega_\tsub{rf}^2}\;\textnormal{,}
  \label{mathieu_a}
\end{equation}
with the applied rf and dc voltages $U_\tsub{rf}$ and $U_\tsub{t}$, each corrected by geometrical factors $\kappa$ of order $1$, the distance $d$ of the electrodes from the trap center, and the secular frequency of the static axial confinement $\omega_z$. The purpose of $U_\tsub{t}$ is to break the rotational symmetry in a controlled way and thereby define the radial principal axes. \\

The solution to (\ref{Mathieu}), to lowest order in $q_i$ and $a_i$, is \cite{Major2005}
\begin{equation}
  u_i(t)=\left(u_{0,i}+u_{1,i}\cos(\omega_it)\right)\left(1+\frac{q_i}{2}\cos(\Omega_\tsub{rf}t+\varphi_i)\right)
  \label{ionmotioneqn}
\end{equation}
\begin{equation}
  \textnormal{with}\quad\omega_i=\frac{\Omega_\tsub{rf}}{2}\sqrt{a_i+\frac{q_i^2}{2}}\;\textnormal{,}
\end{equation}
where $u_{0,i}\propto E_{\tsub{dc},i}$ is a displacement of the equilibrium position from the rf node due to any residual static electric field. Equation (\ref{ionmotioneqn}) contains motion at two frequencies: the secular motion at $\omega_i$ is an oscillation in the harmonic trapping potential with the temperature-dependent amplitude $u_{1,i}$, whereas the motion at $\Omega_\tsub{rf}$ is referred to as micromotion. It consists of an unavoidable ``intrinsic'' contribution from the periodic displacement due to the secular motion (IMM), and the so called ``excess micromotion'' (EMM), which occurs whenever there is a residual rf electric field $E_\tsub{rf}$ at the trapping potential minimum.\\

Under the assumption that the radial confinement is determined entirely by the ponderomotive potential \mbox{($a_i\ll q_i^2$)}, the kinetic energy of the micromotion is equal to the radial potential energy. \cite{Wineland1987} This doubles the kinetic energy from thermal excitation for each direction with rf confinement. Assuming an equal temperature $T$ for secular motion in all three dimensions, the total thermal kinetic energy in a linear trap is therefore
\begin{equation}
  \langle E_\tsub{kin}^\tsub{(th)}\rangle=\frac{5}{2}k_BT\;\textnormal{,}
\end{equation}
where $k_B$ denotes the Boltzmann constant.\\

Residual rf electric fields can occur due to a displacement $u_{0,i}$ of the potential minimum from the rf node or due to a phase shift between the rf electrodes. \cite{Berkeland1998} In an ideal linear trap, micromotion only occurs in the radial plane, as there is no axial component of the rf electric field. Real traps, however, break the translational symmetry due to segmentation and machining tolerances. This can give rise to an axial component of $E_\tsub{rf}$, which cannot easily be compensated and should therefore be minimized by design. This is of particular concern when multiple ions are to be stored with low micromotion. \cite{Herschbach2012,Pyka2014}\\

The EMM velocity $\vec{v}_\tsub{emm}$ can be derived by integrating the equation of motion
\begin{equation}
  m\dot{\vec{v}}_\tsub{emm}=Q\vec{E}_\tsub{rf}\cos(\Omega_\tsub{rf}t)\;\textnormal{,}
\end{equation}
and the corresponding mean kinetic energy is
\begin{equation}
  \langle E_\tsub{kin}^\tsub{(emm)}\rangle=\frac{m}{2}\langle v^2\rangle=\frac{1}{m}\left(\frac{Q}{2\Omega_\tsub{rf}}E_\tsub{rf}\right)^2\;\textnormal{.}
\end{equation}

The motion of the ion gives rise to a 2nd-order Doppler shift of
\begin{equation}
  \label{2ndorderDoppler}
  \left\langle\frac{\Delta\nu_\tsub{D2}}{\nu}\right\rangle=-\frac{\langle E_\tsub{kin}\rangle}{mc^2}=-\left(\frac{Q}{2mc\Omega_\tsub{rf}}E_\tsub{rf}\right)^2-\frac{5k_BT}{2mc^2}\;\textnormal{,}
\end{equation}
where $E_\tsub{kin}$ denotes the total kinetic energy. In addition, the ion experiences a 2nd-order Stark shift due to the non-vanishing mean-squared electric field $\langle E^2\rangle$, which can be derived by averaging the square of the quadrupole field
\begin{equation}
  E^2_q(u_i(t))\approx\left((\nabla_i \vec{E}_q)u_i(t)\right)^2=\left(\frac{q_i\Omega_\tsub{rf}^2m}{2Q}u_i(t)\right)^2
\end{equation}
over the classical ion trajectory (\ref{ionmotioneqn}) for all directions of rf confinement.
The sign and magnitude of the resulting shift
\begin{equation}
  \left\langle\frac{\Delta\nu_\tsub{S}}{\nu}\right\rangle=\sigma_\tsub{S}\left\langle E^2\right\rangle=\sigma_\tsub{S}\left(\frac{1}{2}E_\tsub{rf}^2+\frac{m\Omega_\tsub{rf}^2}{Q^2}k_BT\right)\;\textnormal{,} 
\end{equation}
depend on the differential static polarizability $\sigma_\tsub{S}$ of the involved states. \cite{Berkeland1998} For clock transitions in which the Stark effect increases the frequency, the proper choice of trap drive frequency allows a cancellation of both contributions. \cite{Madej2012,Dube2014} In species with a low differential static polarizability of the clock states, such as In${}^+$ and Al${}^+$, \cite{Safronova2011} the Stark shift due to a given rf electric field amplitude is typically more than an order of magnitude below the corresponding 2nd-order Doppler shift. Although we therefore quantify our results in terms of the 2nd-order Doppler shift, the Stark shift can be derived from the same data, given sufficiently precise knowledge of $\sigma_\tsub{S}$.\\

\section{\label{emm_section}Detection of excess micromotion}
In the frame of an ion undergoing periodic motion with the velocity $\vec{v}=\vec{v}_0\cos(\Omega_\tsub{rf}t+\varphi)$, the 1st-order Doppler shift can be expressed as a laser phase modulation with modulation index $\beta=\vec{k}\cdot\vec{v}_0/\Omega_\tsub{rf}$, where $\vec{k}$ is the wave vector of the laser. For $\beta\ll1$, the resulting spectrum can be approximated by \cite{Riehle2004}
\begin{equation}
  \label{lasersidebands}
  \begin{split}
    E(\omega)&\propto J_0(\beta)\delta(\omega-\omega_L)\\
    &+J_1(\beta)\left(\delta(\omega-\omega_L-\Omega_\tsub{rf})-\delta(\omega-\omega_L+\Omega_\tsub{rf})\right)\;\textnormal{,}
  \end{split}
\end{equation}
where $\omega_L$ is the frequency of the laser in the laboratory reference frame and $J_i$ are the Bessel functions of the first kind. The two quantitative methods evaluated in this work both determine $\beta$, as it is a direct measure of the micromotion amplitude along $\vec{k}$. The residual rf electric field can be derived from $\beta$ via
\begin{equation}
  E_\tsub{rf}=\frac{m\Omega_\tsub{rf}^2}{kQ}\beta\;\textnormal{.}
\end{equation}

\subsection{\label{sb_basics}The resolved sideband method}
The resolved sideband method makes use of a transition with a linewidth of $\Gamma\ll\Omega_\tsub{rf}$. In this case, the micromotion component parallel to $\vec{k}$ can be determined from a measurement of the relative strengths of carrier and sideband transitions in Eq.~(\ref{lasersidebands}):
\begin{equation}
  \frac{\Omega_{\pm1}}{\Omega_0}=\frac{J_1(\beta)}{J_0(\beta)}\approx\frac{\beta}{2}\;\textnormal{,}
\end{equation}
where $\Omega_0$ and $\Omega_{\pm1}$ denote the Rabi frequency of the carrier transition and 1st-order micromotion sideband, respectively.\\

Depending on $\Gamma$ and the available laser power, the Rabi frequencies can be determined either from coherent population evolution or from steady-state excitation rates. In the latter case, decoherence must be well understood, e.g. solely due to excited state decay, in order to quantitatively determine $\Omega_{0,\pm1}$. Sufficiently broad transitions allow a direct detection of the steady-state fluorescence. \cite{Peik1999}\\

\begin{figure}
  \centerline{\includegraphics[width=.4\textwidth]{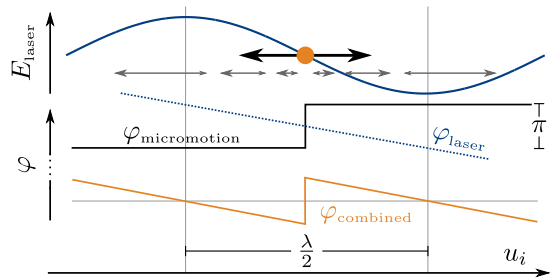}}
  \caption{\label{sideband_imm_sampling}Spatial phase dependence of the IMM signal. Top: ion undergoing secular motion along the $\vec{k}$ vector of the probing laser. The induced micromotion is assumed to be parallel to the secular motion (thin arrows). Center: while the laser phase changes linearly with position, the phase of the micromotion jumps by $\pi$ as the rf node is crossed. Bottom: if the amplitude $u_{1,i}$ of the secular motion becomes comparable to $\lambda/4$, contributions from either side of the rf node no longer cancel entirely.}
\end{figure}

According to Eq.~(\ref{ionmotioneqn}), in the absence of EMM, the ion motion contains only Fourier components at $\omega_i$ and $\Omega_\tsub{rf}\pm\omega_i$. However, the phase modulation of the laser electric field is described by the factor $\exp(i\vec{k}\cdot\vec{u}(t))$, which gives rise to a frequency component at $\Omega_\tsub{rf}$ due to its nonlinearity. The amplitude of this component scales with $q_iu_{1,i}^2$ and is therefore temperature-dependent. Figure \ref{sideband_imm_sampling} illustrates a more intuitive approach: If it is not deep within the Lamb-Dicke regime, \cite{Wineland1998} the ion samples a considerable range of laser phases during its secular motion. This phase change can prevent the cancellation of the out-of-phase IMM contributions from either half of the secular period. Geometrical considerations for the two-dimensional case are discussed in appendix \ref{sbappendix}, along with a quantum-mechanical derivation which shows that for a Fock state with $n$ phonons, the signal can become as big as
\begin{equation}
  \left\vert\frac{\Omega_{\pm1}}{\Omega_0}\right\vert=\frac{q_i}{4}\eta_i^2\left(2n+1\right) + \mathcal{O}(\eta_i^4)\;\textnormal{,}
  \label{imm_sideband}
\end{equation}
where $\eta_i=k\sqrt{\hbar/(2m\omega_i)}$ is the Lamb-Dicke parameter of the mode parallel to $\vec{k}$. The signal for a thermal state can be derived from this expression as an ensemble average of Rabi oscillations.

\subsection{\label{pc_theory_analytic}The photon-correlation method}
The photon-correlation method determines the modulation index $\beta$ using a transition that violates the resolved sideband condition $\Gamma\ll\Omega_\tsub{rf}$. It is based on the correlation between the trap drive voltage and the fluorescence modulation due to EMM, \cite{Diedrich1987,Hoeffges1997} which is observed by recording the time delay between each detected photon and the succeeding rf zero crossing. A histogram of the observed delays resembles the time-inverted distribution of photons within one rf period. It has the form
\begin{equation}
  \label{pcsignal}
  S(t)=S_0+\Delta S \cos (\Omega_\tsub{rf} t - \varphi)\;\textnormal{,}
\end{equation}
where $S_0$ is proportional to the mean fluorescence, the observation time and the bin size, and the normalized modulation amplitude $\Delta S/S_0$ and the phase $\varphi$ are used to determine the amplitude and phase of the EMM.\\

The method has been described in detail by Berkeland et al. \cite{Berkeland1998} for the case when $\Gamma\gg\Omega_\tsub{rf}$, i.e. assuming that the scattering rate is in steady state at every instance of the rf cycle. However, with typical rf drive frequencies on the order of a few $\unit[10]{MHz}$, this assumption cannot be made for e.g. the $S\leftrightarrow P$ transitions in alkali-like ions. We derive in the following the signals obtained by the photon-correlation method in a way that is also valid in the regime $\Gamma\approx\Omega_\tsub{rf}$.\\

\begin{figure}
  \centerline{\includegraphics[width=.45\textwidth]{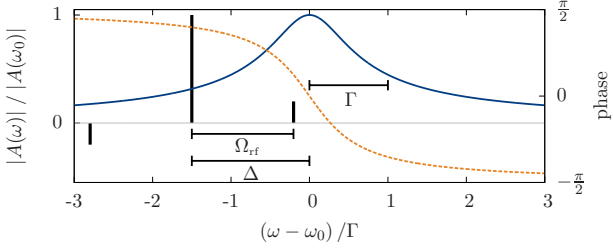}}
  \caption{\label{fmspec_scheme}Illustration of the model used to derive the photon-correlation signal. The atomic transition is described as a classical damped oscillator (see Eq.~(\ref{atom_classical_oscillator})), the spectral response of which is shown by the solid (amplitude) and dashed (phase) lines. In its frame of reference, micromotion causes an apparent phase modulation of the laser, which produces sidebands (vertical bars, see Eq.~(\ref{lasersidebands})). The quantities $\Gamma$,$\Omega_\tsub{rf}$ and $\Delta$ are indicated in units of $\Gamma$.}
\end{figure}

In the limit of low intensity $I\ll I_\tsub{sat}$, the atomic transition can be described as a classical damped harmonic oscillator with resonance (angular) frequency $\omega_0$ and damping rate $\Gamma$. Its frequency response to an excitation at frequency $\omega$ is \cite{Riehle2004}
\begin{equation}
  A(\omega-\omega_0)=\frac{1}{2}\frac{\frac{\Gamma}{2}-i(\omega-\omega_0)}{(\omega-\omega_0)^2+(\frac{\Gamma}{2})^2}\;\textnormal{,}
  \label{atom_classical_oscillator}
\end{equation}
as shown in Fig.~\ref{fmspec_scheme}. The detected fluorescence is proportional to the power in this oscillator when excited with the spectrum (\ref{lasersidebands}), which in the time domain consists of three contributions:
\begin{widetext}
  \begin{equation}
    \label{fmspecresult}
    \begin{split}
      S(\Delta,t)=\left\vert\int_{-\infty}^\infty A(\omega-\omega_0)E(\omega)e^{i\omega t}d\omega\right\vert^2&\propto\underbrace{J^2_0(\beta)\vert A(\Delta)\vert^2+J^2_1(\beta)\left(\vert A(\Delta+\Omega_\tsub{rf})\vert^2+\vert A(\Delta-\Omega_\tsub{rf})\vert^2\right)}_{S_0,\textnormal{``dc component''}}\\
      &+\underbrace{2J_0(\beta)J_1(\beta)\left\vert A^*(\Delta)A(\Delta+\Omega_\tsub{rf})-A(\Delta)A^*(\Delta-\Omega_\tsub{rf})\right\vert}_{\Delta S,\textnormal{``rf component''}}\cos\left(\Omega_\tsub{rf}t+\varphi\right)\\
      &+2J^2_1(\beta)\vert A(\Delta+\Omega_\tsub{rf})A^*(\Delta-\Omega_\tsub{rf})\vert\cos (2\Omega_\tsub{rf}t + \varphi^\prime)\\
    \end{split}
  \end{equation}
  with phase
  \begin{equation}
    \label{fmspecresultphase}
    \varphi=\textnormal{arg}\left(A^*(\Delta)A(\Delta+\Omega_\tsub{rf})-A(\Delta)A^*(\Delta-\Omega_\tsub{rf})\right)\;\textnormal{.}
  \end{equation}
\end{widetext}
Here, $\Delta=\omega_L-\omega_0$ is the detuning of the laser (carrier) frequency from the atomic resonance.\\

\begin{figure}
  \centerline{\includegraphics[width=.5\textwidth]{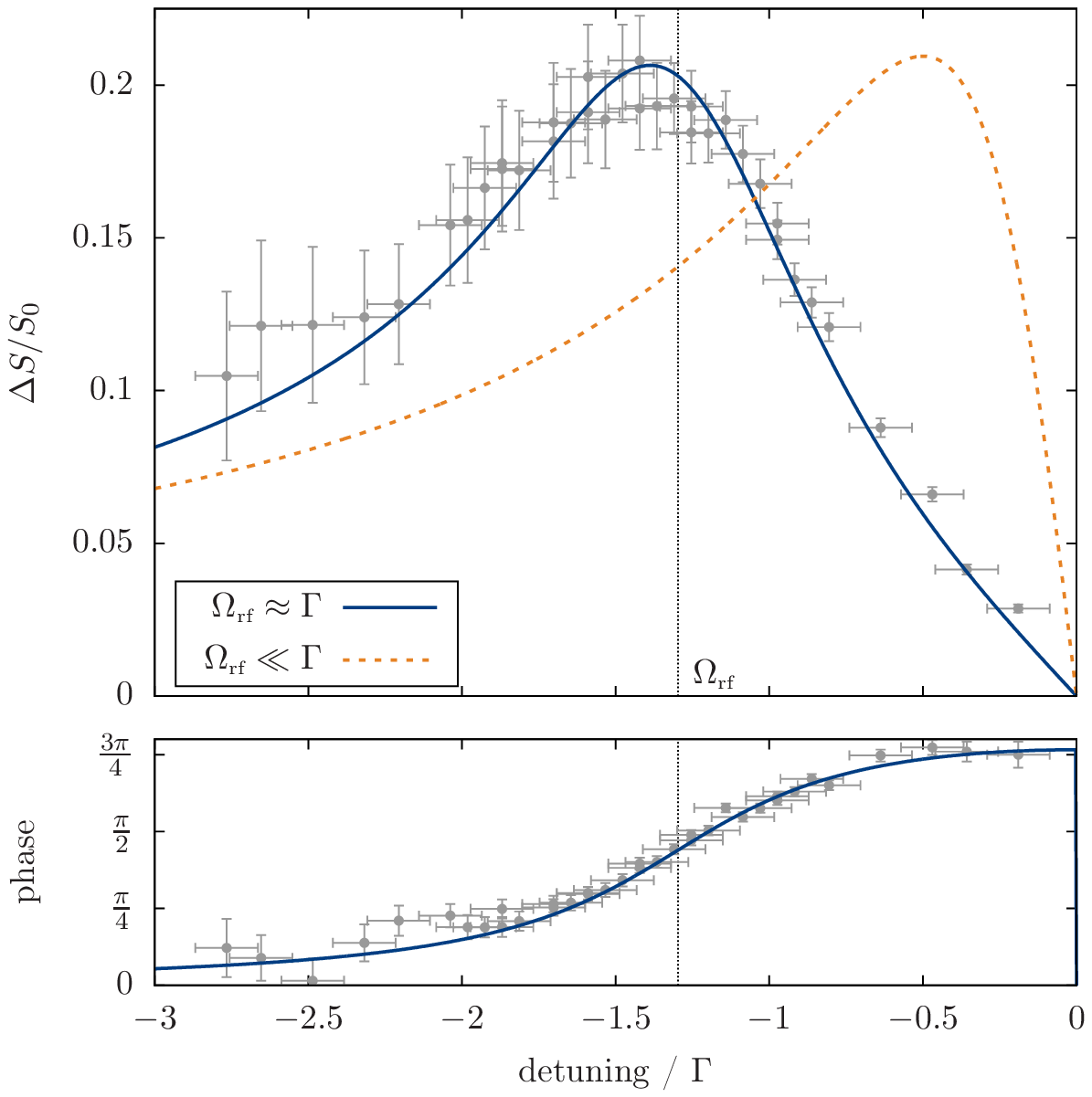}}
  \caption{\label{fmspecresultplot}Photon-correlation signal for a fixed micromotion amplitude ($\beta=0.085$) and varied laser detuning in the limit $I\ll I_\tsub{sat}$. For the solid curve, \mbox{$\Gamma=2\pi\times\unit[19.6]{MHz}$} and \mbox{$\Omega_\tsub{rf}=2\pi\times\unit[25.42]{MHz}$} (indicated by the vertical line). Experimental values confirm the expected behavior (see section \ref{experimental_results}). The dashed curve shows the result derived by Berkeland et al. \cite{Berkeland1998} in the limit $\Omega_\tsub{rf}\ll\Gamma$. The models agree when that assumption is valid.}
\end{figure}

The first two terms in (\ref{fmspecresult}) correspond to the offset $S_0$ and amplitude $\Delta S$ of the observed signal (\ref{pcsignal}), whereas the term that oscillates at $2 \Omega_\tsub{rf}$ is of order $J_1^2(\beta)$ and can be neglected for $\beta\ll1$. This result reflects that the process is analogous to performing frequency-modulation spectroscopy on the atomic transition. \cite{Bjorklund1980}\\

Equation (\ref{fmspecresult}) shows that for $\beta\ll1$, the quantity $\Delta S/S_0\propto J_1(\beta)/J_0(\beta)$ is a direct measure of the EMM amplitude. The proportionality factor depends on the usually well-known parameters $\Gamma$ and $\Omega_\tsub{rf}$, as well as the laser detuning $\Delta$, as shown in Fig.~\ref{fmspecresultplot}. While our model agrees with the one derived by Berkeland et al. \cite{Berkeland1998} in the limit $\Omega_\tsub{rf}\ll\Gamma$, there is a significant deviation when the frequency scales become comparable. This is due to the fact that the phase difference in the atomic response to carrier and sidebands cannot be neglected, and a first order approximation of the line shape over a range of $2\Omega_\tsub{rf}$ fails in this regime (see Fig.~\ref{fmspec_scheme}). It is also apparent that the highest sensitivity is no longer attained at a detuning of $-\Gamma/2$, but rather close to $-\Omega_\tsub{rf}$. Note that the clear maximum at a known detuning provides a simple way to experimentally determine the resonance frequency, which cannot be measured directly on a cooling transition. \cite{Pruttivarasin2014} Its exact amplitude and position depend on the laser intensity if saturation cannot be neglected. This dependence is discussed in appendix \ref{pcappendix}, along with further corrections to Eq.~(\ref{fmspecresult}).

The phase information obtained with the signal is a major advantage of the photon-correlation method: the separation of in-phase and out-of-phase components in measurements with two nonparallel beams allows a full determination of the amplitude and orientation of the micromotion within the common plane. The phase information also allows a distinction between excess micromotion due to a displacement and due to an rf phase shift, since the phase of the former changes by $\pi$ when the ion is moved across the rf node.

\subsection{\label{heating_basics}Micromotion minimization using parametric excitation}
The third method we evaluate does not measure micromotion directly, but detects a displacement $u_{0,i}$ from the minimum of the ponderomotive potential. When a modulation of the rf voltage is applied at $2\omega_\tsub{sec}/n$, with $n\in\mathbb{N}$, the secular motion is excited parametrically at a rate proportional to the displacement. \cite{Vedel1990,Ibaraki2011} Unlike the other two methods, parametric excitation is not able to measure micromotion due to a phase shift between rf electrodes. Besides the simple experimental implementation, a major advantage of this method is that it uses the Doppler shift due to secular motion, rather than micromotion, to generate a signal. In principle, this allows three-dimensional micromotion compensation with a single laser beam, as long as it has projections onto all principal axes of the trap.\\

\begin{figure}
  \centerline{\includegraphics[width=.5\textwidth]{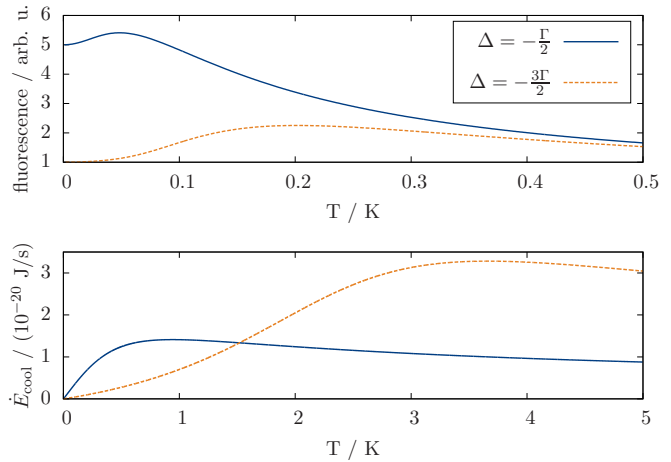}}
  \caption{\label{fluorescenceVsT}Quantities to consider when choosing the laser cooling parameters in parametric excitation measurements: Fluorescence (top) and Doppler cooling rate (bottom) as a function of temperature for laser detunings of $-\Gamma/2$ (solid lines) and $-3\Gamma/2$ (dashed lines) at an intensity of $I=0.1I_\tsub{sat}$.}
\end{figure}

The signals observed for this method are changes in the amount of fluorescence as the modulation frequency is swept over the resonance. Both increases and decreases of the fluorescence have been observed, depending on the choice of parameters. \cite{Ibaraki2011,Narayanan2011,Tanaka2012} An increase in fluorescence typically occurs for low deviations from Doppler temperature, whereas fluorescence decreases when the temperature is increased further. In order to get a monotonic dependence of the signal with respect to EMM, it is therefore necessary to operate in the regime of fluorescence increase. The corresponding temperature range can be extended by detuning the cooling laser further from resonance, as shown in Fig.~\ref{fluorescenceVsT} (top), which shows the fluorescence based on the value of a Voigt profile at detunings of $-\Gamma/2$ and $-3\Gamma/2$, respectively. Since both the heating and Doppler cooling rates depend on temperature, experimental parameters must be chosen such that equilibrium is eventually reached, to avoid runaway heating. As the sensitivity of the method increases with equilibrium temperature, there is a trade-off between resolution and robustness, which can require several iterations while adjusting parameters as the minimum is approached.\\

The Doppler cooling rate can be estimated by integrating the energy loss due to the momentum imparted by absorbed photons \cite{Wineland1979} over one period of secular motion:
\begin{align}
  \label{dcrate}
  &\dot{E}_\tsub{cool}=\frac{I}{I_\tsub{sat}}\hbar kv_0\frac{\Gamma}{4\pi}\int_0^{2\pi}\frac{\sin(t^\prime)}{1+\left(\frac{2\left(\Delta-kv_0\sin(t^\prime)\right)}{\Gamma}\right)^2}dt^\prime\\&\textnormal{with}\;v_0=\sqrt{\frac{k_BT}{m}}\;\textnormal{,}\nonumber
\end{align}
where $k$ is the projection of $\vec{k}$ onto the principal axis of the mode, and we have assumed $I\ll I_\tsub{sat}$. Equation (\ref{dcrate}) neglects heating due to the recoil from emitted photons, which has negligible influence for temperatures far above the Doppler limit. Figure \ref{fluorescenceVsT} (bottom) shows the numerically calculated values of Eq.~(\ref{dcrate}) as a function of $T$.\\

The heating rate due to parametric excitation has been derived by Savard et al. \cite{Savard1997} as
\begin{equation}
  \langle\dot{E}_\tsub{exc}\rangle=\frac{m\omega_{i,0}^2}{2\tau}\int_0^\tau\dot{\varepsilon}(t^\prime)u_i^2(t^\prime)dt^\prime\;\textnormal{,}
\end{equation}
where the secular frequency is modulated as $\omega_i(t)=\omega_{i,0}\sqrt{1+\varepsilon(t)}$ and $\tau$ is short compared to the heating rate. To give an example, for the experimental parameters described in section \ref{parametric_heating} and $E_\tsub{rf}=\unitfrac[100]{V}{m}$, the initial heating rate at $T=\unit[0.5]{mK}$ is on the order of $\unitfrac[10^{-21}]{J}{s}$. In the experiments, the fluorescence begins to drop for $E_\tsub{rf}>\unitfrac[100]{V}{m}$, indicating that the equilibrium temperature exceeds $\unit[200]{mK}$ at this point.

\subsection{\label{experimental_results}Experimental comparison}

\begin{figure}
  \centerline{\includegraphics[width=.45\textwidth]{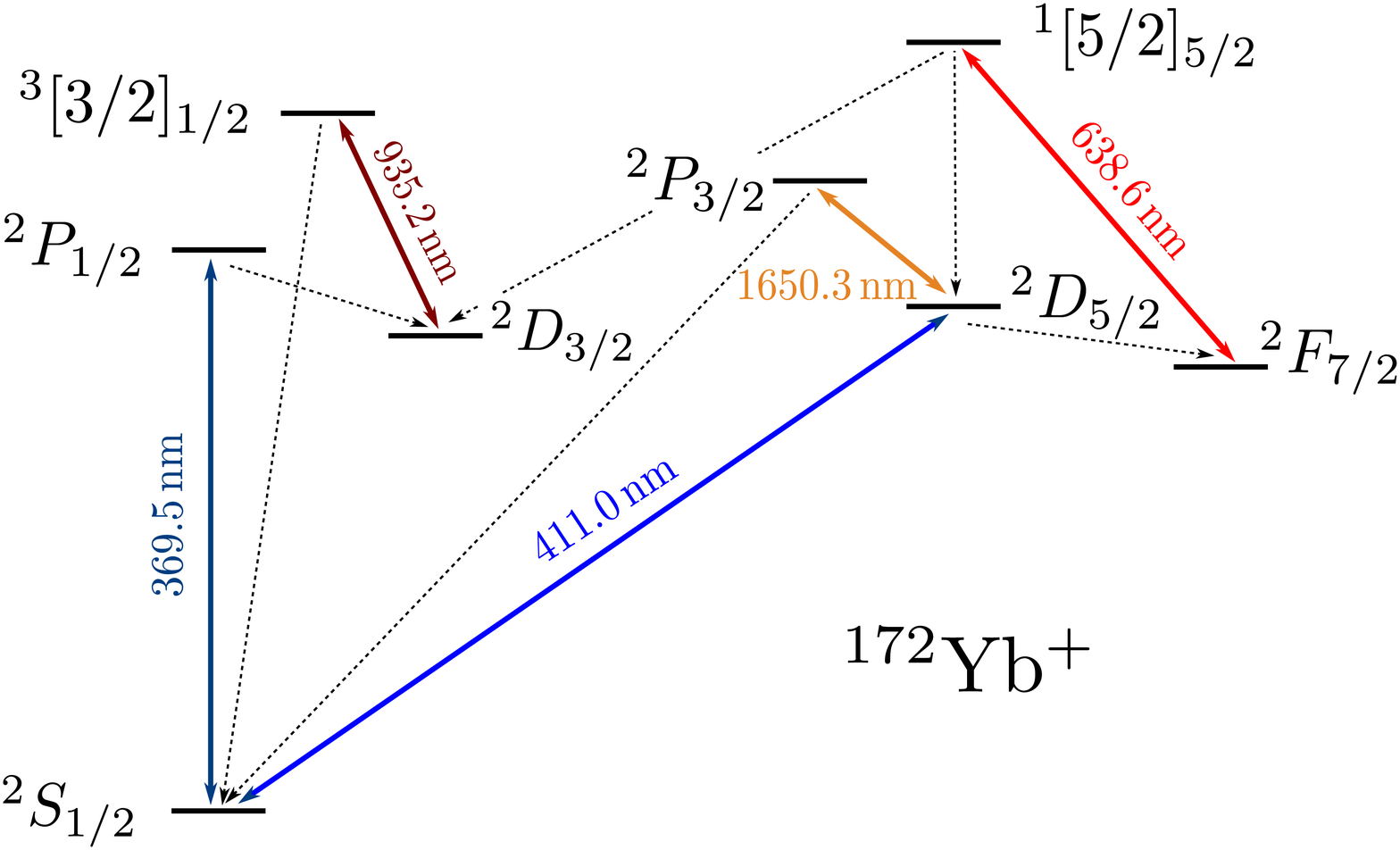}}
  \caption{\label{term_scheme}Reduced term scheme of ${}^{172}$Yb${}^+$. The electric dipole transition at $\unit[369.5]{nm}$ is used for Doppler cooling, state detection, and photon-correlation measurements. Sideband measurements are performed on the $\unit[411]{nm}$ electric quadrupole transition. For ground state cooling, the ${}^2$D${}_{5/2}$ state is depleted via the $\unit[1650]{nm}$ transition to the short-lived ${}^2$P${}_{3/2}$ state.}
\end{figure}

All of the methods are tested using a single ${}^{172}$Yb${}^+$ ion. Figure \ref{term_scheme} shows the relevant levels and transitions. Light at $\unit[369.5]{nm}$ can be applied from two horizontal directions (in the $x$-$z$ plane as defined in Fig.~\ref{trap_schematic}, at $\unit[25]{{}^\circ}$ and $\unit[155]{{}^\circ}$ with respect to the $z$ axis), and vertically. The collected fluorescence is split and imaged onto a photomultiplier tube (PMT) and an electron multiplying CCD camera (EMCCD). The setup is described in more detail in a previous publication. \cite{Pyka2014} We use an rf drive frequency of $\Omega_\tsub{rf}=2\pi\times\unit[25.42]{MHz}$. In order to find the resolution limits and to test the agreement of quantitative evaluations, interleaved measurements comparing two methods each are performed while scanning the compensation voltage $U_\tsub{ec}$ (see Fig.~\ref{trap_schematic}). This shifts the ion almost entirely (to within $\unit[2]{{}^\circ}$) along the $x$ direction, inducing micromotion along $y$. In the case of sideband and photon-correlation measurements, this component is then detected using a vertical beam. An additional horizontal cooling beam is applied during vertical photon-correlation measurements to prevent axial heating of the ion. Its power is two orders of magnitude below that of the vertical beam, such that the influence on the fluorescence signal can be neglected. Before each scan, micromotion along $x$ is measured with photon-correlation measurements using the horizontal beams and minimized by adjusting the compensation voltage $U_\tsub{tc}$. Axial micromotion can be inferred from the same data and minimized by shifting the axial ion position. After an initial adjustment, no significant increase in axial micromotion (within an $E_\tsub{rf,z}$ uncertainty of $\unitfrac[40]{V}{m}$) has been observed over the course of four months.

\subsubsection{\label{pc_vs_sb}Photon-correlation and sideband methods}
\begin{figure}
  \centerline{\includegraphics[width=.5\textwidth]{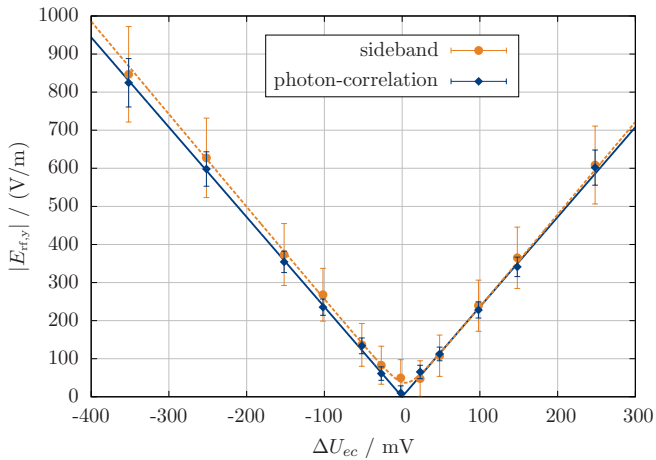}}
  \caption{\label{interleaved_corr_411}Interleaved measurement using the photon-correlation and sideband methods (at $\unit[0.5]{mK}$) as the ion is shifted radially through the trap center. The error bars of the photon correlation data are due to shot noise and the systematic uncertainties discussed in appendix \ref{pcappendix}. The error bars of the sideband method assume $\unit[10]{\%}$ fit uncertainty of the Rabi frequencies. The function fitted to the sideband data takes into account the offset due to IMM discussed in section \ref{sb_basics}.}
\end{figure}

For photon-correlation measurements, the \mbox{${}^2$S${}_{1/2}$} $\leftrightarrow$ \mbox{${}^2$P${}_{1/2}$} transition (\mbox{$\lambda=\unit[369.5]{nm}$}, \mbox{$\Gamma=2\pi\times\unit[19.6]{MHz}$}) is used. A time-to-amplitude converter measures the delay between each photon detected by the PMT and the following zero crossing of the rf voltage. Measured intervals are then binned and counted by a multi-channel analyzer. The amplitude and phase of the micromotion are obtained by fitting Eq.~(\ref{pcsignal}) to the resulting signal. A conservative integration time of $\unit[90]{s}$ is used for each measurement, although much shorter intervals could be used without significantly deteriorating the resolution, as shown in section \ref{clocksection}.\\

Sideband measurements are performed on the \mbox{${}^2$S${}_{1/2}\leftrightarrow {}^2$D${}_{5/2}$} transition (\mbox{$\lambda=\unit[411]{nm}$}, \mbox{$\Gamma=2\pi\times\unit[23]{Hz}$}), addressed by an ultra-stable laser. \cite{Keller2014} For state detection, the fluorescence on the $\unit[369.5]{nm}$ transition is observed with the EMCCD camera. All measurements are performed close to the Doppler limit of $\unit[0.5]{mK}$ at radial secular frequencies around $2\pi\times\unit[470]{kHz}$. A spectroscopy pulse of varied duration is applied (200 experimental cycles each), and a fit with Rabi oscillations of a thermal ensemble is used to extract the ground state Rabi frequencies for both the carrier and micromotion sideband.\\

Figure \ref{interleaved_corr_411} shows the result of a comparison between photon-correlation and sideband measurements. The fitted slopes of $E_\tsub{rf}$ with respect to $U_\tsub{ec}$ are \mbox{$s_\tsub{pc}=\unit[(2360\pm30)]{(V/m)/V}$} and \mbox{$s_\tsub{sb}=\unit[(2437\pm29)]{(V/m)/V}$} for the photon-correlation and sideband method, respectively, showing a residual mismatch of less than $\unit[4]{\%}$. The optimum values of $U_\tsub{ec}$ determined by the fits agree to within $\unit[4.2]{mV}$, corresponding to a difference in $E_\tsub{rf}$ of $\unitfrac[(10 \pm 4)]{V}{m}$. We currently have no explanation for this discrepancy, which corresponds to a 2nd-order Doppler shift of $3\times10^{-21}$. As explained in appendix \ref{pcappendix}, radiation pressure during the photon-correlation measurements can be excluded as a reason.\\

While the photon-correlation result at the minimum is compatible with zero micromotion, there is a statistically significant offset of about $\unitfrac[37]{V}{m}$ when using the sideband method at a temperature of $T=\unit[0.5]{mK}$. Measurements with varied Doppler cooling parameters show a monotonic dependence of $\Omega_{+1}$ on temperature. When the ion is cooled to the motional ground state in the directions of rf confinement, no excitation on the sideband is observed. We therefore attribute this signal to IMM, as described in section \ref{sb_basics}. The observed amplitude is about one third of the maximum value as predicted by Eq.~(\ref{imm_sideband}). As explained in appendix \ref{sbappendix}, the signal is observed in our geometry due to the lifted degeneracy of the radial secular frequencies. The determination of the optimal compensation voltage is not affected, since it can be interpolated from measurements at higher displacements, where excess micromotion dominates.

\subsubsection{\label{parametric_heating}Parametric excitation}
\begin{figure}
  \centerline{\includegraphics[width=.5\textwidth]{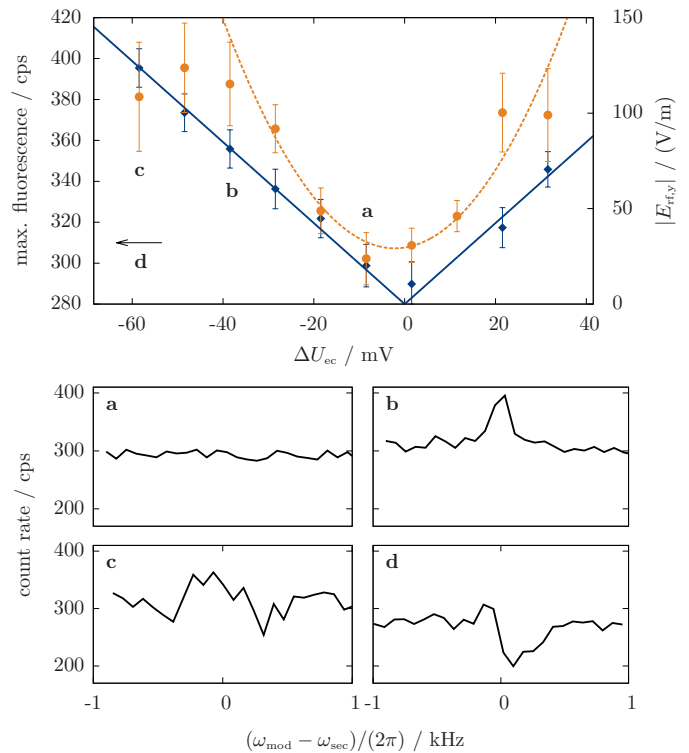}}
  \caption{\label{interleaved_corr_heating}Parametric excitation signals. Top: fluorescence at the parametric resonance as the ion is shifted radially through the trap (orange circles; mean and standard deviation of 10 sweeps each). A parabola is fitted to the six innermost data points to determine the position of the minimum. Intermittent photon-correlation measurements are used to quantify the amount of excess micromotion (blue diamonds). Bottom: examples of signals obtained in different micromotion amplitude regimes. Low amounts of micromotion lead to an increase of fluorescence, whereas strong micromotion produces a fluorescence decrease. Labels \textit{a} to \textit{d} indicate the corresponding data points in the upper graph.}
\end{figure}

The parameters for parametric excitation measurements are chosen as a compromise between sensitivity and robustness. As shown in section \ref{heating_basics}, a higher detuning of the cooling laser from resonance widens the temperature range for which both fluorescence and cooling rate increase with increasing temperature. The measurements are performed at a detuning $\Delta=-3\Gamma/2$ and intensity $I/I_\tsub{sat}\approx0.01$. The parametric resonance is excited by an amplitude modulation of the rf voltage, which modulates the secular frequency as
\begin{equation}
  \omega_\tsub{sec}(t)=\omega_\tsub{sec,0}\left(1+m_0\sin(\omega_\tsub{mod}t)\right)\;\textnormal{,}
\end{equation}
with $\omega_\tsub{sec,0}=2\pi\times\unit[500]{kHz}$, and modulation depth $m_0=0.04$. The modulation frequency $\omega_\tsub{mod}$ is swept over a range of $\unit[2\pi\times2]{kHz}$ around $\omega_\tsub{sec,0}$ during $\unit[30]{s}$.
Figure \ref{interleaved_corr_heating} shows a parametric heating measurement as the ion is moved along the $x$ direction using the compensation voltage $U_\tsub{ec}$. Each point is derived from the maximum fluorescence of the average of 10 modulation frequency sweeps, examples of which are shown in the bottom part of the figure. Interleaved photon-correlation measurements are performed to quantify the amount of micromotion along $y$. It should be noted that the two measurements are sensitive to displacements along different axes: while the amount of vertical micromotion determined by photon-correlation is proportional to the $x$ displacement, the strength of parametric excitation depends on the displacement along the respective principal axis, which is at about $-\unit[48]{{}^\circ}$ from the $x$ axis. This dependence has been verified by displacing the ion by the same amount in different directions. To ensure that the minima in Fig.~\ref{interleaved_corr_heating} coincide, the $y$ displacement of the ion was minimized beforehand using $U_\tsub{tc}$ and photon-correlation measurements with the horizontal beams. The fitted minima agree to within $\Delta U_\tsub{ec}=\unit[2]{mV}$, corresponding to a mismatch of $E_\tsub{rf}=\unitfrac[(4.6\pm3.7)]{V}{m}$. The sensitivity of the parametric excitation method can be increased by lowering the trapping potential in order to increase the displacement due to a given residual static field $E_\tsub{dc}$.\\

Parametric excitation could also be used for the detection of axial micromotion. However, the contribution of the rf field to the axial trapping potential strongly depends on the electrode geometry and is ideally negligible by design, in particular when compared to the static axial potential. Exciting axial motion would therefore require a much stronger modulation of the rf voltage.\\

\subsubsection{Resolution limits}
Table \ref{exp_value_comparison_table} summarizes the resolution limit we observe for a single measurement with each method and gives an overview of other published values.\\

\begingroup
\squeezetable
\begin{table*}
  \begin{ruledtabular}
    \begin{tabular}{llllllllll}
      \tabularnewline[-0.3cm]
      \textbf{method}&\textbf{species}&$\unit[\lambda]{/nm}$&$\unit[\Omega_\tsub{rf}]{/MHz}$&$\unit[\omega]{/MHz}$&$\beta\times10^3$&$\unit[E_\tsub{rf}]{/(V/m)}$&$\unit[E_{dc}]{/(V/m)}$&$\vert\Delta\nu_\tsub{D2}/\nu\vert\times10^{20}$&\textbf{ref.}\tabularnewline\toprule\tabularnewline[-0.2cm]
      sideband&${}^{40}$Ca${}^+$&$729$&$23.5$&$3.4$&$\mathbf{4\pm1}$&$4.2\pm1.1$&$0.42\pm0.1$&$1.3\pm0.7$&\cite{Chwalla2009}\tabularnewline
      &${}^{40}$Ca${}^+$&$397$&$55$&$4$&$0\pm60$&$0\pm197$&$\mathbf{0\pm10}$&$0\pm530$&\cite{Doret2012}\tabularnewline
      sideband ($T_\tsub{Doppler}$)&${}^{172}$Yb${}^+$&$411$&$25.4$&$0.47$&$17\pm16$&$50\pm48$&$0.6\pm0.6$&$8.5\pm16$&this work\tabularnewline\hline\tabularnewline[-0.2cm]
      photon-correlation&${}^{199}$Hg${}^+$&$194$&$8.6$&$0.065$&&&&$\mathbf{0\pm20}$&\cite{Berkeland1998}\tabularnewline
      &${}^{88}$Sr${}^+$&$422$&$23.2$&&$60$&$78$&&$95$&\cite{Wilpers2012}\tabularnewline
      &${}^{172}$Yb${}^+$&$370$&$25.4$&$0.47$&$2.7\pm 2.7$&$7.1\pm7.2$&$0.09\pm0.09$&$0.17\pm0.35$&this work\tabularnewline\hline\tabularnewline[-0.2cm]
      parametric excitation&${}^{40}$Ca${}^+$&&$15$&$1.2$&&$22$&$1.22$&$90$&\cite{Narayanan2011}\tabularnewline
      &${}^{40}$Ca${}^+$&&$14$&$1.5$&&$0\pm64$&$\mathbf{0\pm6}$&$0\pm860$&\cite{Tanaka2012}\tabularnewline
      &${}^{172}$Yb${}^+$&&$25.4$&$0.47$&&$0\pm23$&$0\pm0.3$&$0\pm1.8$&this work\tabularnewline\hline\tabularnewline[-0.2cm]
      neutral atom probe&${}^{87}$Rb${}^+$&&$4.17$&$0.35$&&$0\pm0.34$&$\mathbf{0\pm0.02}$&$0\pm0.057$&\cite{Haerter2013}\tabularnewline\hline\tabularnewline[-0.2cm]
      high finesse cavity&${}^{138}$Ba${}^+$&$493$&$5.3$&$1.2$&$89\pm3$&$11.1\pm0.3$&$1.73\pm0.05$&$15.1\pm0.9$&\cite{Chuah2013}\tabularnewline\hline\tabularnewline[-0.2cm]
      repumper Doppler&${}^{40}$Ca${}^+$&$866$&$25.8$&$3.1$&$0\pm8$&$0\pm12$&$\mathbf{0\pm1}$&$0\pm8.6$&\cite{Allcock2010}\tabularnewline\hline\tabularnewline[-0.2cm]
      ion trajectory vs. $U_\tsub{rf}$&${}^{172}$Yb${}^+$&&$13.2$&$0.133$&&$0\pm19$&$\mathbf{0\pm0.09}$&$0\pm4.65$&\cite{Gloger2015}\tabularnewline
    \end{tabular}
  \end{ruledtabular}
  \caption{\label{exp_value_comparison_table}Comparison of the sensitivity achieved in a single measurement with each technique and overview of other published micromotion measurements. For the methods not discussed in this article, the lowest published values are shown. Bold figures indicate the quantity given in the reference. Note that $\beta$ and $\vert\Delta\nu_\tsub{D2}/\nu\vert$ quantify EMM independent of the mass. Factors of $2\pi$ have been omitted in all $\Omega_\tsub{rf}$ and $\omega$ values for clarity. Note that the sideband value quoted for this work is measured at Doppler temperature and therefore limited by the IMM signal described in section \ref{sb_basics}.}
\end{table*}
\endgroup

In sideband measurements, the resolution is limited by the lowest observable Rabi frequency on the sideband and the available laser power. Currently, we can achieve $\Omega_0=2\pi\times\unit[140]{kHz}$, limited by the output power of the second harmonic generation for the $\unit[411]{nm}$ light. The sideband Rabi frequency resolution is $\Omega_{\pm1}=2\pi\times\unit[500]{Hz}$, limited by magnetic field fluctuations on the order of $\unit[30]{nT}$ between experimental cycles. The lowest resolvable modulation index due to these technical limitations is $\beta=0.007$, which corresponds to $E_\tsub{rf}=\unitfrac[21]{V}{m}$. At a temperature of $\unit[0.5]{mK}$ however, the minimum sideband excitation we observe is limited by IMM, as described in section \ref{pc_vs_sb}. The fit uncertainty in obtaining $\Omega_{0,\pm1}$ from measurements using an ion at the Doppler limit is about $\unit[10]{\%}$. Taking all these contributions into account, we experimentally observe a minimum of $E_\tsub{rf}=\unitfrac[(50\pm48)]{V}{m}$ for an ion at the Doppler temperature. In the ground state, the expected IMM contribution corresponds to less than $\unitfrac[1]{V}{m}$ and Rabi frequencies can be determined more precisely.\\

The photon correlation method is limited by shot noise, as discussed in section \ref{clocksection}. It contributes about $\unitfrac[5]{V}{m}$ to the uncertainty in $E_\tsub{rf}$ for an integration time of $\unit[90]{s}$. Taking into account the additional uncertainty contributions listed in appendix \ref{pcappendix}, we experimentally observe a minimum of $\beta=(2.7\pm2.7)\times10^{-3}$, which corresponds to $E_\tsub{rf}=\unitfrac[(7.1\pm7.2)]{V}{m}$.\\

For the parametric excitation method, the statistical uncertainty of the fluorescence peak is given, as shown in Fig.~\ref{interleaved_corr_heating}. It is dominated by photon shot noise and corresponds to $E_\tsub{rf}=\unitfrac[23]{V}{m}$.

\subsection{\label{clocksection}Applicability for optical clock operation}
During clock operation, varying external electric fields need to be compensated at regular intervals. If the frequency feedback to the clock laser is interrupted while the new compensation voltage values are determined, the required uncertainty needs to be achieved as quickly as possible. \cite{Barwood2015} In the following, we discuss the time consumption of micromotion compensation with a single ion. In a multi-ion clock, an initial evaluation over the full extent of the ion crystal is necessary in order to determine the frequency shifts of the individual ions. However, for crystals with a small extension compared to the distance from the electrodes, it is reasonable to approximate fluctuating fields as homogeneous. Drifts in EMM can therefore be compensated using measurements on a single ion as well. In a segmented trap, the Coulomb crystal can meanwhile be stored in a separate segment.\\

As shown in section \ref{pc_vs_sb}, the quantitative dependence of micromotion on the radial displacement can be mapped precisely, see e.g. Fig.~\ref{interleaved_corr_411}. A single measurement per dimension is therefore sufficient to determine the minimum. A controlled displacement is needed to resolve the sign ambiguity when sideband or parametric heating measurements are used. For photon-correlation measurements, this is not necessary due to the contained phase information.\\

\begin{figure}
  \centerline{\includegraphics[width=.4\textwidth]{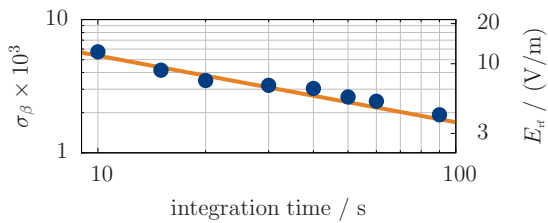}}
  \caption{\label{min_deltaS_vs_t}Uncertainty of the photon-correlation method due to shot noise as a function of integration time. The modulation index $\beta$ is calculated for $\Omega_\tsub{rf}=2\pi\times\unit[25.4]{MHz}$ and $\lambda=\unit[369.5]{nm}$. The fitted uncertainty is $1.7\times10^{-2}/\sqrt{t}$ for a count rate of $\unit[3600]{cps}$. All other contributions to uncertainty are typically more than an order of magnitude lower.}
\end{figure}

In a photon-correlation measurement, most of the uncertainty contributions are relative and therefore negligible due to the fact that the micromotion amplitude is already low. In this case the dominating contribution, by more than an order of magnitude, is the statistical uncertainty of $\Delta S$. In order to evaluate its dependence on integration time, we perform a series of photon-correlation measurements with a straylight signal. Figure \ref{min_deltaS_vs_t} shows the standard deviation of fit results for $\Delta S$ of 50 measurements each as a function of the integration time $t$, expressed as the uncertainty in $\beta$ and $E_\tsub{rf}$ for our experimental parameters. The function $\sigma_\beta(t)=\sigma_\beta(\unit[1]{s})/\sqrt{t}$ fits well to the data, indicating that the uncertainty is solely due to photon shot noise. Since straylight equally contributes to this uncertainty, a high signal to noise ratio is helpful. In our case, $\unit[10]{s}$ of measurement per dimension would therefore be sufficient to reduce the respective uncertainty in the 2nd-order Doppler shift to $1\times10^{-20}$. Since the signals can be derived from the fluorescence during Doppler cooling, there is no fundamental need to interrupt clock operation at all when applying this method. For this purpose, cooling beams need to be applied from alternating directions to allow three-dimensional compensation. With sufficiently low trap heating rates, strongly attenuated auxiliary beams can be used to ensure cooling of all motional modes, as described for the vertical measurements in section \ref{experimental_results}.\\

In a sideband measurement without ground state cooling, the acquisition of a single data point (200 cycles) takes at least $\unit[2]{s}$, limited by the time required for state detection and repumping. Total time consumption per dimension is therefore on the order of $\unit[10]{s}$. Since the optimal compensation voltage is extrapolated from a measurement with added EMM, it is affected less by the IMM contribution and decoherence than measurements with minimized EMM.\\

Parametric heating measurements need to be performed within the regime of monotonic dependence of the fluorescence peak on the micromotion amplitude if the minimum is to be deduced from a single data point. The choice of parameters therefore requires a trade-off between robustness and resolution. As the fluorescence value depends on the equilibrium temperature in the presence of heating, laser cooling parameters need to be well controlled in order to achieve reproducible values. On a timescale of hours, we observe a secular frequency instability on the order of $\unit[100]{Hz}$, which is comparable to the width of the parametric resonance. This requires either an active stabilization of $U_\tsub{rf}$ or a scan of the modulation frequency. Assuming an averaging time of $\unit[10]{s}$ per modulation frequency, total time consumption is therefore on the order of $10$ to $\unit[100]{s}$.\\

\section{\label{summary}Summary and conclusions}
\begingroup
\squeezetable
\begin{table*}
  \begin{ruledtabular}
    \begin{tabular}{p{.1\textwidth}p{.32\textwidth}p{.20\textwidth}p{.09\textwidth}p{.11\textwidth}}
      \tabularnewline[-0.3cm]
      \textbf{method}&\textbf{advantages}&\textbf{limitations}&\textbf{resolution}&\textbf{time per point}\tabularnewline\toprule\tabularnewline[-0.2cm]
      sideband&fast acquisition\newline measures ratio $\Omega_1/\Omega_0$\newline $\rightarrow$ common mode effects cancel&resolution limited by\newline -- decoherence\newline -- laser power&$3\times10^{-20}$\newline $\sigma_\beta=0.007$&$<\unit[10]{s}$\tabularnewline\hline\tabularnewline[-0.2cm]
      photon-correlation&contains phase information\newline$\rightarrow$ $E_\tsub{rf}$ direction can be inferred,\newline \phantom{$\rightarrow$} rf phase shift / displacement distinguishable\newline applicable continuously during clock operation&resolution limited by\newline -- photon shot noise&$3.5\times10^{-21}$\newline $\sigma_\beta=0.0027$&$\unit[10]{s}$ to $\unit[100]{s}$\newline (no dead time)\tabularnewline\hline\tabularnewline[-0.2cm]
      parametric excitation&simple implementation\newline no 3D laser access required&no quantitative evaluation\newline insensitive to rf phase shift&$2\times10^{-20}$&$\unit[10]{s}$ to $\unit[100]{s}$\tabularnewline
    \end{tabular}
  \end{ruledtabular}
  \caption{\label{method_summary_table}Summary of the methods evaluated in this work. Resolutions achieved under our experimental conditions are given in terms of the second-order Doppler shift. For the quantitative methods, the corresponding uncertainty in the modulation index $\sigma_\beta$ is indicated.}
\end{table*}
\endgroup

Precision spectroscopy with trapped ions requires an accurate determination of motional frequency shifts. In this article, we experimentally compare three commonly used methods for the compensation and characterization of excess micromotion and show that each of them is capable of ensuring a 2nd-order Doppler shift well below $10^{-19}$, independent of the ion mass. Table \ref{method_summary_table} summarizes the advantages and limitations of these methods as well as their experimentally obtained single data point resolution and time consumption. A new model allows us to quantitatively evaluate photon-correlation measurements in the common regime where the transition linewidth is on the order of the rf drive frequency. The obtained micromotion amplitudes agree well with those obtained in resolved sideband measurements. While ground state cooling is not necessary for the resolved sideband method, a temperature-dependent offset on the order of $10^{-19}$ limits the resolution if the ion is not deep within the Lamb-Dicke regime. It is shown to be due to sampled intrinsic micromotion. The optimal compensation voltages determined from interleaved scans agree to within less than $\unit[5]{mV}$, which corresponds to a residual 2nd-order Doppler shift uncertainty of $3.4\times10^{-21}$ (photon-correlation/sideband) and $1\times10^{-21}$ (photon-correlation/parametric heating). The corresponding 2nd-order Stark shift due to excess micromotion can be deduced from the same measurements, given sufficiently precise knowledge of the differential static polarizability. In the case of In${}^+$ and Al${}^+$, this contribution is negligible. \cite{Safronova2011}

\begin{acknowledgments}
  The authors thank I.~D. Leroux and D.~J. Wineland for fruitful discussions, and E. Peik and P.~O. Schmidt for helpful comments on the manuscript. This work was funded by DFG through QUEST and by the EU through SIB04 -- Ion Clock. The EMRP is jointly funded by the EMRP participating countries within EURAMET and the European Union.
\end{acknowledgments}

\begin{appendix}

  \section{\label{sbappendix}Sideband method}
  In this appendix, we derive the amount of IMM sampled by the sideband method, as given by Eq.~(\ref{imm_sideband}). For simplicity, we first consider only motion in one dimension, assuming that the secular motion, the resulting micromotion, and the $\vec{k}$ vector are parallel. Furthermore, we neglect the pulsation of the harmonic oscillator wavefunctions at $\Omega_\tsub{rf}$. \cite{Leibfried2003}\\
  
  In the adiabatic approximation of Eq.~(\ref{ionmotioneqn}), micromotion can be described as a position-dependent phase modulation with amplitude $k\hat{x}q/2$, where $\hat{x}=x_0(\hat{a}+\hat{a}^\dagger)$ is the position operator, and the rf node defines the origin. A laser field detuned by $\Omega_\tsub{rf}$ appears in the moving frame of reference as
  \begin{equation}
    \vec{E}(t)=\vec{E}_0e^{i(\omega_0+\Omega_\tsub{rf})t}e^{ik\hat{x}}e^{i\frac{1}{2}k\hat{x}q\cos(\Omega_\tsub{rf}t)}\;\textnormal{.}
    \label{imm_qm_derivation_1}
  \end{equation}
  Expanding the rightmost exponential as a series of sidebands weighted with the respective Bessel functions yields
  \begin{equation}
    \vec{E}(t)=\vec{E}_0e^{i(\omega_0+\Omega_\tsub{rf})t}e^{ik\hat{x}}\sum_{n=-\infty}^\infty i^nJ_n\left(\frac{k\hat{x}q}{2}\right)e^{in\Omega_\tsub{rf}t}\;\textnormal{.}
  \end{equation}
  Neglecting all terms detuned by $\geq\Omega_\tsub{rf}$ reduces the expression to
  \begin{equation}
    \vec{E}(t)=\vec{E}_0e^{i\omega_0t}e^{ik\hat{x}}(-i)J_{-1}\left(\frac{k\hat{x}q}{2}\right)\;\textnormal{.}
  \end{equation}
  The transition matrix element between internal states $\vert g\rangle$ and $\vert e\rangle$ in the motional state $\vert n\rangle$ is
  \begin{equation}
    \left\langle e, n\left\vert \vec{d} \cdot \vec{E}(t) \right\vert g, n\right\rangle=\frac{\hbar\Omega_0}{2}\left\langle n\left\vert e^{ik\hat{x}}(-i)J_{-1}\left(\frac{k\hat{x}q}{2}\right)\right\vert n\right\rangle\;\textnormal{,}
  \end{equation}
  where $\vec{d}$ denotes the transition dipole element. The overlap integral of the motional wavefunction evaluates to
  \begin{equation}
    \left\langle n\left\vert e^{ik\hat{x}}(-i)J_{-1}\left(\frac{k\hat{x}q}{2}\right)\right\vert n\right\rangle=\frac{q}{4}\eta^2\left(2n+1\right)+\mathcal{O}\left(\eta^4\right)\;\textnormal{,}
    \label{imm_qm_derivation_5}
  \end{equation}
  with the Lamb-Dicke parameter $\eta=kx_0$. It is straightforward to include EMM in the above derivation by adding an offset to $\hat{x}$ in the rightmost exponential in (\ref{imm_qm_derivation_1}). This results in an imaginary second term in (\ref{imm_qm_derivation_5}) with an amplitude of $\langle n\vert e^{ik\hat{x}}\vert n\rangle \beta/2 + \mathcal{O}(q^3)$, as expected for the EMM sideband. The $\pi/2$ relative phase between the terms shows that the two components cannot interfere and therefore the overall minimum still corresponds to compensated EMM.\\
  
  \begin{figure}
    \centerline{\includegraphics[width=.4\textwidth]{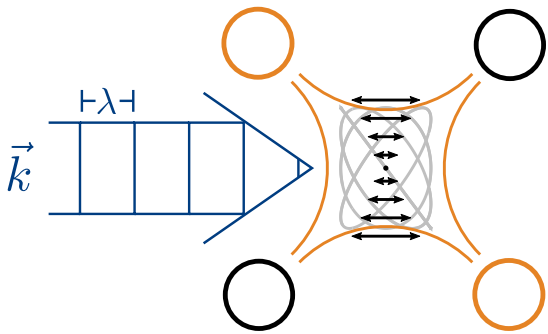}}
    \caption{\label{sideband_imm_sampling_2D}Illustration of the sampled IMM in resolved sideband measurements for a typical geometry (not to scale). The signal depends on the amount of correlation between secular motion (gray lines) and micromotion (black arrows) in the radial trajectories of the ion when projected onto $\vec{k}$, averaged over the duration of the pulse.}
  \end{figure}
  
  Since the $\hat{x}^2$ term that produces the IMM signal requires correlated micromotion and secular motion components along $\vec{k}$, its amplitude is maximized in the geometry assumed above. In typical setups, optical access is such that $\vec{k}$ must be aligned almost perpendicular to the quadrupole field lines, as depicted in Fig.~\ref{sideband_imm_sampling_2D}. The relative amplitude of the signal at Doppler temperature can then be calculated from the temporal average of the product of secular motion and micromotion in the classical radial trajectory of the ion, each projected onto $\vec{k}$. For degenerate principal axes, it vanishes completely due to symmetry considerations. If the degeneracy is lifted, the amplitude depends on the pulse duration compared to the difference in secular frequencies, the amount of splitting, and the orientation of the principal axes.

  \section{\label{pcappendix}Photon-correlation method}
  This appendix lists corrections to Eq.~(\ref{fmspecresult}) and experimental sources of uncertainty for the excess micromotion amplitude. For orientation, example values are calculated for the ${}^2S_{1/2}\leftrightarrow {}^{2}P_{1/2}$ transition in ${}^{172}$Yb${}^+$ at $\unit[369.5]{nm}$, which has a natural linewidth of $\Gamma=2\pi\times\unit[19.6]{MHz}$. For the trap drive frequency, we assume $\Omega_\tsub{rf}=2\pi\times\unit[25.4]{MHz}$.
  
  \subsection{RF pickup}
  
\begin{figure}
    \centerline{\includegraphics[width=.5\textwidth]{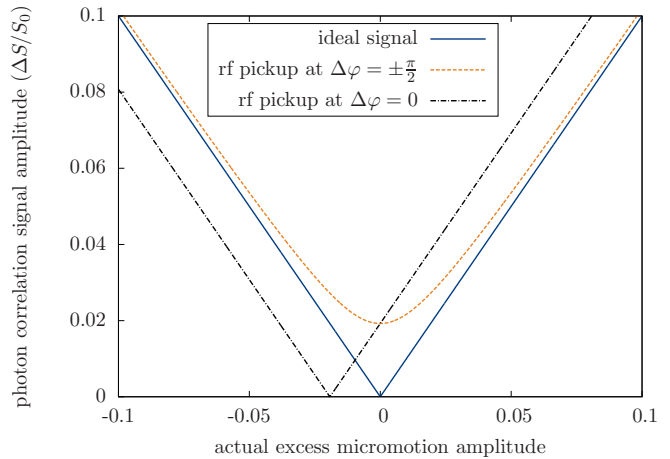}}
    \caption{\label{pickup}Influence of rf pickup on the photon-correlation signal. Depending on the relative phase of the disturbance with respect to actual excess micromotion, it can either shift the minimum position (dash-dotted line) or prevent the minimum amplitude from being measured (dashed line).}
  \end{figure}

  Since there is usually a high amount of power at $\Omega_\tsub{rf}$ present in the trap drive electronics, it is difficult to prevent parasitic coupling to nearby signals. As with any measurement that relies on the demodulation of a signal at a given frequency, photon-correlation measurements are very sensitive to this kind of disturbance. The influence on the measurement depends on the relative phase between the real signal and the pickup and is shown in Fig.~\ref{pickup} for the two limiting cases: If the phase difference is exactly $\pi/2$, the two contributions add up quadratically, leading to an overestimation of the amount of residual micromotion when minimized. Due to the fact that the phase of the actual micromotion shifts by $\pi$ when the rf node is crossed, an in-phase pickup signal will shift the observed position of the minimum.
  If the problem cannot be solved through shielding of the respective components, it can be eliminated in data processing. In our setup, \cite{Pyka2014} we observe a modulation of the PMT detection efficiency with $\Omega_\tsub{rf}$, which remains at a constant amplitude and phase for weeks as long as no part of either the rf circuit or PMT readout electronics is moved. Its typical influence is an error in $E_\tsub{rf}$ on the order of $\unitfrac[20]{V}{m}$. This kind of disturbance can be removed from the raw data after its relative amplitude and phase have been determined by running the measurement cycle with only straylight (if necessary for a longer integration time than the actual measurements).
  
  \subsection{\label{pc_theory_numerical}Saturation}
  
  \begin{figure}
    \centerline{\includegraphics[width=.5\textwidth]{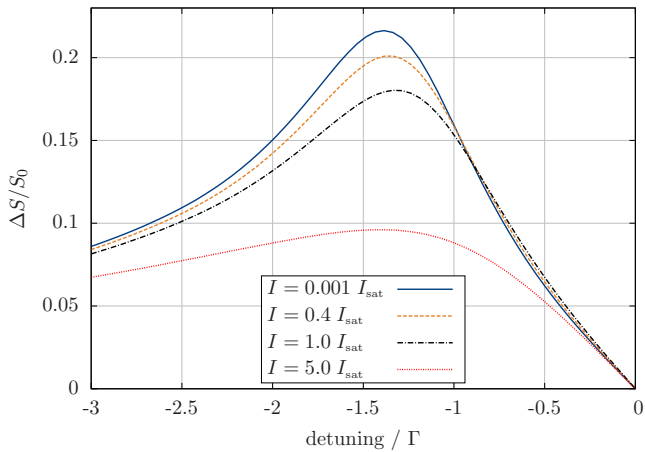}}
    \caption{\label{meresults}Numerically calculated photon-correlation signal amplitudes as a function of laser detuning when taking saturation into account. The other relevant parameters are $\Omega_\tsub{rf}=2\pi\times\unit[25.42]{MHz}$, $\Gamma=2\pi\times\unit[19.6]{MHz}$, and $\beta=0.085$.}
  \end{figure}
  While the treatment in section \ref{pc_theory_analytic} has the advantage of providing an analytic expression, the classical approach cannot take saturation into account. Experimentally, it is not desirable to measure at low intensities, since longer integration times are necessary to obtain a given signal to noise ratio. In order to include saturation in our model, we numerically integrate the master equation of a laser interacting with a two-level atom in the presence of a sinusoidally varied detuning. \cite{Johansson2013} After a sufficient amount of time to avoid the influence of transients, a fit to the temporal evolution of the excited state population extracts the signal amplitude $\Delta S/S_0$ and phase $\varphi$. Fig.~\ref{meresults} shows the frequency dependence of $\Delta S/S_0$ for a fixed amount of excess micromotion at various intensities. The low intensity result matches the analytic expression (\ref{fmspecresult}).
  
  The dependence of signal amplitude on saturation gives rise to an experimental uncertainty. As a compromise between sensitivity and photon shot noise, we use $I=0.4\times I_\tsub{sat}$ for our measurements. A first order approximation of the numerical results shows that
  \begin{equation}
    \delta\left(\frac{\Delta S}{S_0}\right)_{I}\approx\frac{\Delta S}{S_0}\times(-0.175)\times\delta\left(\frac{I}{I_\tsub{sat}}\right)\;\textnormal{.}
  \end{equation}
  As the laser power is actively stabilized, the biggest contribution to $\delta(I/I_\tsub{sat})$ is the uncertainty in the saturation power. At $\unit[25]{\%}$, it gives rise to a $\unit[1.8]{\%}$ relative uncertainty in $\Delta S/S_0$.
  
  \subsection{Zeeman shifts}
  To avoid the existence of dark states, multiple Zeeman components of the transition need to be addressed. If their frequencies differ by a considerable fraction of the transition linewidth, the influence on the signals shown in Fig.~\ref{meresults} cannot be neglected. This can be incorporated into the model as an ensemble average. For the measurements in section \ref{experimental_results}, we apply a bias field of $\unit[100]{\mu T}$ along $\vec{k}$ and use linearly polarized light, such that only $\sigma_\pm$ transitions are driven. The line is therefore split into two equal components that shift in opposite directions by the same amount. Figure \ref{emmvsdetuning_broadening} shows the effect on the photon-correlation signal amplitude and the detuning at which the maximum occurs, as a function of this shift.

  \subsection{Other line broadening effects}
\begin{figure}
    \centerline{\includegraphics[width=.5\textwidth]{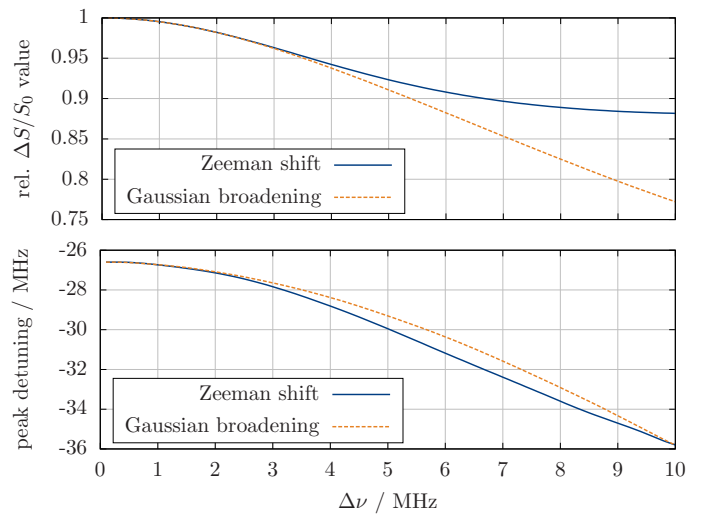}}
    \caption{\label{emmvsdetuning_broadening}Influence of line broadening effects on the photon-correlation signal. Zeeman broadening assumes two components, shifted with opposite signs by the amount indicated as $\Delta\nu$ relative to the unperturbed resonance.  For Gaussian broadening, $\Delta\nu$ refers to the Gaussian $\sigma$. Top: relative maximum value of $\Delta S / S_0$ compared to the unbroadened case. Bottom: detuning for which the sensitivity is maximized.}
  \end{figure}

  The transition is broadened further by the 1st-order Doppler shift and laser frequency noise. We approximate these effects by an ensemble average over a Gaussian distribution of detunings. Figure \ref{emmvsdetuning_broadening} shows the result as a function of the standard deviation of this distribution.
  
  \subsection{Radiation pressure}
  The continuous illumination during the measurement produces a net force that shifts the equilibrium position of the ion by
  \begin{equation}
    \Delta \vec{x} = \frac{\hbar \vec{k} \Gamma_\tsub{sc}}{m\omega^2}\;\textnormal{,}
  \end{equation}
  where $\Gamma_\tsub{sc}$ is the photon scattering rate, and $\omega$ quantifies the restoring force of the trapping potential in the direction of $\vec{k}$. For our experimental parameters and $\omega=2\pi\times\unit[440]{kHz}$, this amounts to $\Delta x=\unit[2.2]{nm}$. Our choice of $\vec{k}$ ensures that the micromotion induced by this displacement is perpendicular to $\vec{k}$ and therefore does not affect compensation.
  
  \subsection{Further sources of uncertainty}
  
  As shown in section \ref{pc_theory_analytic}, the signal strength for a fixed amount of micromotion depends on laser detuning. This is a 2nd-order effect if the measurement is done at the sensitivity maximum, and in our case it results in a relative uncertainty of
  \begin{equation}
    \delta\left(\frac{\Delta S}{S_0}\right)_{\nu}\approx\frac{\Delta S}{S_0}\times(-4.4\times10^{-3})\unit{MHz^{-2}}\times\delta\nu^2\;\textnormal{.}
  \end{equation}
  After the atomic resonance frequency is determined by maximizing $\Delta S/S_0$ for a fixed amount of micromotion, we stabilize the laser to a wavemeter, \cite{Pyka2014} which in turn is calibrated with an ultra-stable reference laser every $\unit[600]{s}$. This limits the absolute frequency uncertainty to $\pm\unit[2]{MHz}$, which leads to a relative uncertainty contribution of $1.8\%$.\\

  The phase information contained in photon-correlation signals allows coordinate transformations of measurements done with multiple beams when the $\vec{k}$ vectors cannot be aligned with the axes of interest. However, the phase dependence on laser detuning, shown in the lower part of Fig.~\ref{fmspecresultplot}, means that frequency fluctuations increase the uncertainty of the outcome.\\
  
\end{appendix}
\bibliography{emm_paper}

\end{document}